\documentclass[preprint]{aastex}      % 1-col, 1-space

\newcommand{\halpha}{H$\alpha$}
\newcommand{\hbeta}{H$\beta$}

\slugcomment{Accepted: Astrophysical Journal (Letters)}

\shortauthors{Lee, Richer, \& McCall}
\shorttitle{UGC 7636: Stripping In The Virgo Cluster}

\begin{document}

\title{
The Dwarf Irregular Galaxy UGC 7636 Exposed : \\
Stripping At Work In The Virgo Cluster 
}

\author{Henry Lee\altaffilmark{1} }
\affil{Department of Physics and Astronomy, York University,
4700 Keele St., \\ Toronto, Ontario M3J 1P3 Canada}
\email{lee@aries.phys.yorku.ca}

\author{Michael G. Richer\altaffilmark{1,2} }
\affil{Instituto de Astronom\'{\i}a, UNAM, Apartado Postal 
70-264, Ciudad Universitaria, \\ M\'exico D.F., 04510 M\'exico}
\email{richer@astrosen.unam.mx}

\and

\author{Marshall L. McCall}
\affil{Department of Physics and Astronomy, York University,
4700 Keele St., \\ Toronto, Ontario M3J 1P3 Canada}
\email{mccall@aries.phys.yorku.ca}

\altaffiltext{1}{Visiting Astronomer, Canada--France--Hawaii Telescope, 
operated by the National Research Council of Canada, the
Centre National de la Recherche Scientifique de France, 
and the University of Hawaii.}

\altaffiltext{2}{Present address: Observatorio Astron\'omico Nacional,
P.O. Box 439027, San Diego, CA, 92143-9027 USA}

\begin{abstract}
We present the results of optical spectroscopy of a newly discovered
H~II region residing in the H~I gas cloud located between the dwarf
irregular galaxy UGC~7636 and the giant elliptical galaxy NGC~4472 in
the Virgo Cluster. 
By comparing UGC~7636 with dwarf irregular galaxies in the field, we
show that the H~I cloud must have originated from UGC~7636 because 
(1) the oxygen abundance of the cloud agrees with that expected for a
galaxy with the blue luminosity of UGC~7636, and
(2) $M_{H I}/L_B$ for UGC~7636 becomes consistent with the
measured oxygen abundance of the cloud if the H~I mass of the cloud is
added back into UGC~7636.
It is likely that tides from NGC~4472 first loosened the H~I gas,
after which ram--pressure stripping removed the gas from UGC~7636.
\end{abstract}


\section{Introduction} \label{sec-intro}

Outstanding questions regarding the origin and evolution of galaxies
continue to be addressed with studies of dwarf galaxies.
In the hierarchical picture of galaxy formation, dwarf galaxies are
thought to be the initial building blocks which subsequently merged in
the assembly of giant galaxies. 
Comparing properties of dwarf galaxies in the vicinity of the Local
Group with those in nearby clusters may provide some hope
of disentangling the question of ``nature'' versus ``nurture'' 
by evaluating the environmental effects upon galaxy evolution.

The Virgo Cluster is the nearest cluster where one can observe with
relative ease the interaction of galaxies with the intracluster medium.
X--ray observations have shown that the emission in Virgo arises
from the underlying hot intracluster gas 
\citep{edge91,bohringer94,irwin96,schindler99}.
Consequently, star--forming disk galaxies might be affected by the
cluster environment as they pass through the denser regions of the
cluster. 

Virgo spiral galaxies are observed to be H~I--deficient relative to
field spirals, probably as a result of ram--pressure stripping
\citep{cayatte94,kenney99}.
As disk galaxies move at high speeds through the intracluster medium,
ram--pressure stripping occurs when the pressure force exceeds
the gravitational binding force of the disk, resulting in the removal
of the galaxian gas content \citep{gunngott}.
With smaller gravitational potential wells, gas--rich dwarf irregulars
in the Virgo Cluster can satisfy the condition for ram--pressure
stripping \citep{gh89,irwin96}. 
In fact, \citet{vigroux86} interpreted optical and H~I observations of
the Virgo dwarf irregular galaxy \objectname[]{IC~3475 (VCC~1448)} as 
being a consequence of the stripping of its neutral gas content.   

Recent work has indicated that the hot X--ray emitting gas surrounding
the Virgo elliptical galaxy \objectname[]{NGC~4472 (M~49, VCC~1226)} 
\citep{fjt85,fabbiano92,irwin96} may be responsible for removing the
H~I gas from the dwarf irregular galaxy 
\objectname[]{UGC~7636 (VCC~1249)}.
The X--ray emission arises from hot gas associated with NGC~4472
and gas in the intracluster medium \citep{irwin96}. 
It has been suggested that the H~I cloud located
between NGC~4472 and UGC~7636 originated in the dwarf galaxy
\citep{ste87,pt92,henning93,mcnam94,irwin96,lee97}.
For the remainder of the paper, we will refer to the H~I cloud as
STE1, in honour of its discoverers Sancisi, Thonnard, and Ekers.

As part of an ongoing effort to examine the environmental effects
on the chemical evolution of dwarf irregular galaxies in the Virgo
Cluster \citep{lee00}, we undertook an investigation of UGC~7636
and its surrounding environment.
Here, we report the discovery of an H~II region in STE1, along
with a measurement of the oxygen abundance. 
Our measurements show that STE1 once must have resided in UGC~7636 and
that the dwarf galaxy was stripped of its gas.  

Observations and reductions are described in \S~\ref{sec-obs},
the results are discussed in \S~\ref{sec-disc}, and conclusions are
given in \S~\ref{sec-concl}. 
Basic data for UGC~7636 and STE1 are listed in Table~\ref{table-u7636}.
Employing a distance modulus of 18.57 mag for the Large Magellanic Cloud
\citep{panagia98}, we adopt a value of 31.12 mag for the distance modulus
to the Virgo Cluster \citep{ferrarese96}. 

\section{Observations and Reductions} \label{sec-obs}

UGC~7636 and STE1 were observed with the MOS spectrograph at the f/8
focus of the 3.6--m Canada--France--Hawaii Telescope (CFHT) on the
night of 1999 April 11 (UT).
The MOS multi--object imaging grism spectrograph \citep{lefevre94} uses
focal--plane masks created from previously obtained images.
The STIS2 2048$\times$2048 CCD with 21~$\mu$m pixels was used in
combination with the 600 $\ell$/mm grism to provide
wavelength coverage between 3600~\AA\ and 7000~\AA\ 
% (depending on the slit placement) 
at a dispersion of 2.2~\AA\ per pixel.
The spatial scale at the detector was 0\farcs{44} per pixel and the
imaging field was 10\arcmin\ by 10\arcmin.

A five--minute \halpha\ image of a field centred on UGC~7636 was
acquired to determine the locations of possible H~II regions.
Within UGC~7636, faint and diffuse \halpha\ emission was detected, but
no H~II regions were found, confirming the results of \citet{gh89}.
However, at about 2\arcmin\ to the northwest of UGC~7636, a compact
region exhibiting \halpha\ emission was discovered at approximately
the same general position as STE1.
The H~II region coincides with the position of the blue feature
``A'' in \citet{mcnam94} and the young star cluster ``C2'' in Lee et
al. (1997).
The position of the H~II region, which we henceforth call LR1,
is given in Table~\ref{table-u7636}. 

An object mask was constructed online using the YAG laser with
object slits approximately 2\arcsec\ wide and 10\arcsec\ long.
Slits were aligned parallel to the east--west axis.
For the UGC~7636 object mask, two slits were placed lengthwise across
the galaxy and one slit was placed at the position of LR1.
Ne--Ar--Hg arc--lamp spectra were used to calibrate the 
wavelength scale.  
The standard star Feige 67 was observed to calibrate the fluxes.
Illuminated dome flat--field exposures were obtained to remove
pixel--to--pixel sensitivity variations. 

An 817--second spectrum was taken of the entire UGC~7636 field before
increasing fog cover and rising humidity levels forced the termination
of the exposure and closure of the dome shutter. 
Spatial profiles of stars in holes adjacent to the slits displayed a
full width at half maximum of about 2\farcs2.  
A spectrum was successfully recorded from LR1, whereas no signal was
detected from the diffuse \halpha\ emission in UGC~7636. 

% \subsection{Reductions} \label{sec-red}

The spectra were reduced using standard IRAF\footnote{IRAF is
distributed by the National Optical Astronomy Observatories, which is
operated by the Associated Universities for Research in Astronomy,
Inc., under contract to the National Science Foundation.}
routines from the spectroscopy reductions package ``specred.''
Given the small size of the slits, the slit function was assumed to be
constant over the length of each slit.
The large--scale structure in the dispersion direction was removed
from the dome flat exposures, exposing the pixel--to--pixel
variations.
Before final extraction and flux--calibration, cosmic rays in the
spectrum were identified and deleted manually.

The emission--line strengths were measured using locally--developed
software.
Table~\ref{table-data} lists the observed flux ratios, $F$, relative
to \hbeta\ for LR1.
These flux ratios were corrected for underlying Balmer absorption
with equivalent width 2~\AA\ \citep{mrs85} to yield the final flux
ratios, $I$. 
The resulting $I$(\halpha)/$I$(\hbeta) ratio is consistent with the
theoretical value \citep{osterbrock}, indicating no reddening, in
agreement with other observations of Virgo dwarf irregulars
(Lee et al. 2000).
Thus, no reddening correction was applied to our data.
Corrections and analyses were performed with SNAP
(Spreadsheet Nebular Analysis Package, Krawchuk et al. 1997). 
% The listed errors in the ratios include the error in the \hbeta\ flux.
The heliocentric velocity of LR1 computed from Doppler shifts of the
emission lines was found to be consistent with the heliocentric
velocity of STE1 \citep{mcnam94}.

The oxygen abundance was determined using the ``bright-line'' method,
namely the empirical calibration of the emission--line ratio, $R_{23}$ =
([O~II]$\lambda$3727 + [O~III]$\lambda\lambda$4959,5007)/\hbeta\
\citep{pagel79}.
The [N~II]$\lambda$6583/[O~II]$\lambda$3727 ratio was used to break
the degeneracy in the calibration for an appropriate selection
of either the high--abundance or low--abundance branch
\citep{mcgaugh91,mcgaugh97}. 
Despite the uncertain detection of the [N~II] line in the spectrum of
LR1, the maximum possible value of [N~II]/[O~II] is a factor of two
below the threshold value of 0.1, confirming the choice of the
low--abundance branch.  
For a field sample of dwarf irregulars in the vicinity of
the Local Group \citep{rm95}, oxygen abundances derived by this method
agree with those obtained directly from the [O~III]$\lambda$4363 line
to within $\sim$0.2 dex (Lee et al. 2000).

Additional derived properties relevant to the burst of star
formation are listed in Table~\ref{table-data}. 
Using the adopted distance modulus of 31.12 mag, 
the \halpha\ luminosity, the present star formation rate, and
the luminosity of ionizing photons by number were computed
by assuming Case B \citep{osterbrock}.

\section{Discussion} \label{sec-disc}   % DISCUSSION

In Figure~\ref{fig-zmb}, the metallicity measured by the oxygen
abundance, O/H, is plotted versus absolute magnitude for field dwarf
irregulars in the vicinity of the Local Group \citep{rm95}.
The cross locates the measured oxygen abundance of LR1 and the
luminosity of UGC~7636.
Its position falls very close to the locus defined by field dwarf
irregulars. 

A plot of the oxygen abundance versus $M_{H I}/L_B$ in
Figure~\ref{fig-zmh1lb} shows the degree to which UGC~7636 is 
currently deficient in H~I gas compared to field dwarf galaxies
at similar metallicities. 
The cross shows the position of UGC~7636 based upon a measurement of
the remaining H~I content within the galaxy \citep{hoffman87}. 
The two filled symbols indicate where UGC~7636 would lie in
the diagram if gas in STE1 were added to the galaxy.
The filled square and diamond represent the allowable range in H~I gas
mass from single--dish measurements by Sancisi et al. (1987)
and \citet{pt92}, respectively.
Clearly, UGC~7636 would appear as a fairly normal dwarf irregular
if STE1 were a part of the galaxy.

We conclude that STE1 originated in UGC~7636 because the
metallicity of the cloud agrees with what is expected for a galaxy
with the same blue luminosity as UGC~7636, and because $M_{H I}/L_B$
for the galaxy would be consistent with other dwarf irregulars if
the H~I gas in STE1 were added into the galaxy.
There is no evidence of an underlying population of old-- or
intermediate--age stars in STE1 (McNamara et al. 1994; Lee et al. 1997), 
which would have provided the chemical enrichment observed.
The colours of UGC~7636 \citep{gh86,pt96} are fairly normal compared
to those of other dwarf irregular galaxies \citep{pt96,ab98},
indicating that little fading has occurred since gas was removed.
So, STE1 is the former interstellar medium of UGC~7636.  

The unique properties of UGC~7636 and STE1 may stem from a combination of
tidal effects and ram--pressure stripping.
The gravitational binding of the gas was likely first weakened by
tidal forces as UGC~7636 strayed near NGC~4472, which would have
caused the distribution of both stars and gas to become distended as
observed \citep{pt92,mcnam94}. 
Subsequent travel through the hot, dense gas surrounding NGC~4472
\citep{irwin96} would have detached the loosened gas from the stars
via ram--pressure stripping.

We estimate the H~I surface density in STE1 to evaluate whether
there was sufficient H~I for star formation to have occurred.
Converting the peak beam--averaged column density and the mean column
density given in \citet{mcnam94}, the H~I surface density in STE1
is about $\Sigma \sim$ 1--4 $M_{\odot}\;{\rm pc}^{-2}$.
The critical surface density for star formation \citep{kennicutt89}
is given by
$\Sigma_{\rm cr}$ ($M_{\odot}$ pc$^{-2}$) = $0.59 \, \alpha \, V$/$R$,
where $V$ is the maximum rotational velocity in km s$^{-1}$, 
$R$ is the radius in kpc, and $\alpha$ is the stability constant.
Assuming $V \approx$ 40--60 km s$^{-1}$ for 
dwarf irregulars at $M_B \sim -17$ \citep{skillman96},
$R \approx 6$ kpc \citep{mcnam94},
and $\alpha \approx$ 0.3--0.7 \citep{hp96,kennicutt89},
we obtain $\Sigma_{\rm cr} \approx$ 2--4 $M_{\odot}\;{\rm pc}^{-2}$.
Thus, the ratio $\Sigma/\Sigma_{\rm cr}$ is of order unity, satisfying
the threshold criterion for star formation in STE1.

It would be timely with the launch of the Chandra AXAF observatory to
observe in detail the area immediately surrounding LR1 to look for
possible mixing of the H~I gas with the hot, X--ray emitting gas. 
Observing additional dwarf galaxies in the Virgo Cluster
with deep \halpha, optical, and H~I imaging would help to evaluate the
degree to which stripping processes affect the evolution of dwarf
irregular galaxies falling into the central regions of the cluster.

\section{Conclusions} \label{sec-concl}     % CONCLUSIONS

We have presented the results of optical spectroscopy of a newly
discovered H~II region in the H~I gas cloud located between 
the dwarf irregular UGC~7636 and the giant elliptical NGC~4472
in the Virgo Cluster.
A comparison of the global properties for UGC~7636 with dwarf
irregulars in the field shows that the H~I cloud was once
the interstellar medium of UGC~7636.
Tidal forces from NGC~4472 followed by ram--pressure stripping due to
the surrounding hot X--ray emitting gas is likely responsible for
removing the H~I gas from UGC~7636.  

\acknowledgments

H. L. and M. G. R. wish to thank Marshall McCall for financial support, 
and the staff at CFHT for the allocated observing time and their
assistance with the observations. 
M. G. R. also acknowledges support from the Instituto de
Astronom{\'\i}a at UNAM.
M. L. M. would like to thank the Natural Sciences and Engineering 
Research Council of Canada for its continuing support.
Data was accessed as Guest User at the Canadian Astronomy Data Center,
which is operated by the National Research Council, Herzberg Institute of 
Astrophysics, Dominion Astrophysical Observatory. 
This research has made use of the NASA/IPAC Extragalactic Database
(NED) which is operated by the Jet Propulsion Laboratory,
California Institute of Technology, under contract with the
National Aeronautics and Space Administration.

\clearpage	% FIGURES and CAPTIONS

\figcaption[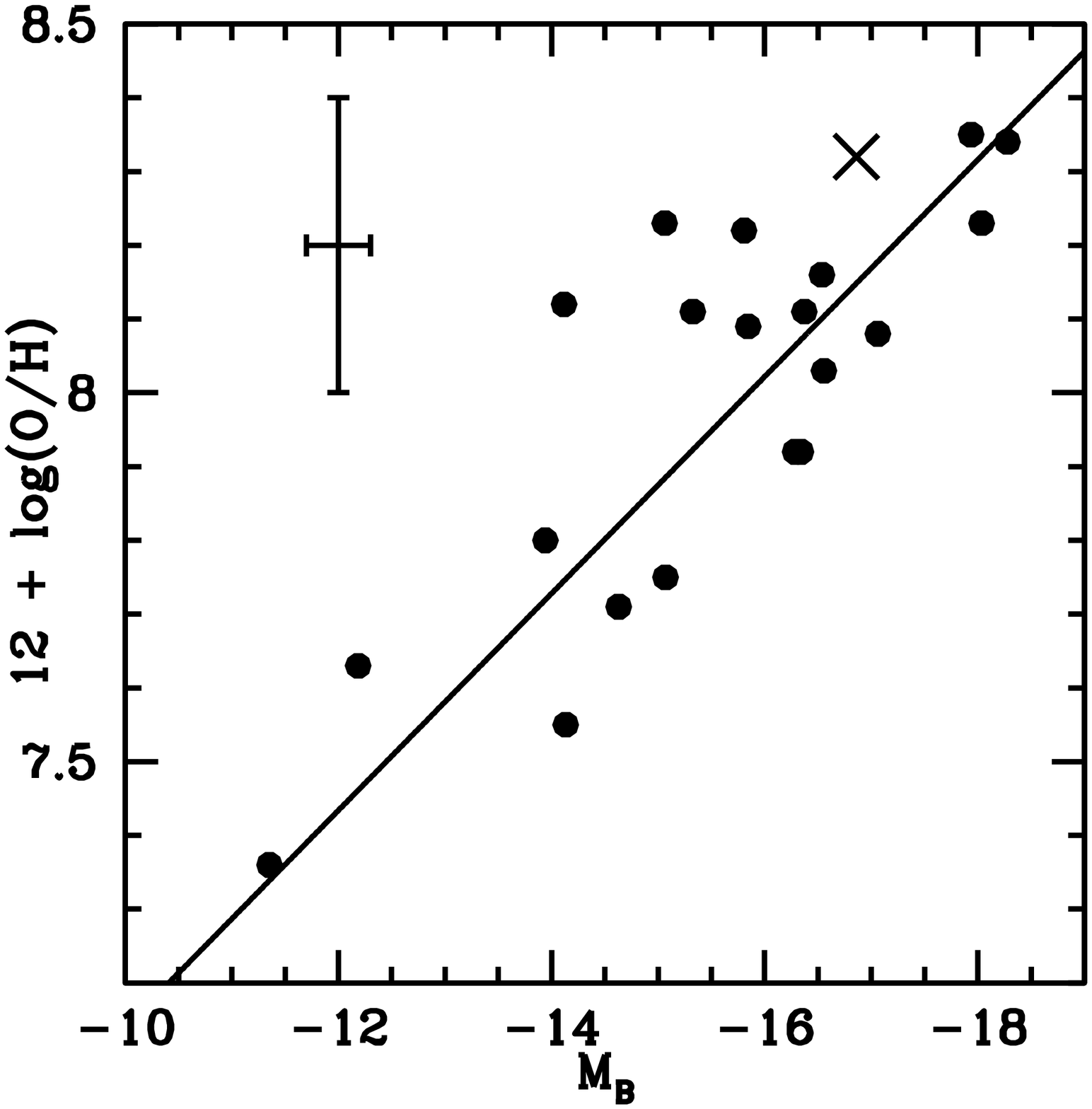]{
Oxygen abundance versus total blue luminosity for dwarf irregular
galaxies.  
The filled circles mark the field dwarfs of Richer \& McCall (1995)
and the solid line is a linear least--squares fit for $M_B < -15$. 
The solar value of the oxygen abundance is 12+log(O/H) = 8.87
(Grevesse, Noels, \& Sauval 1996).
The cross shows where UGC~7636 would lie if its oxygen abundance 
equals that of the newly discovered H~II region, LR1, inside the H~I
cloud STE1. 
The error bar indicates conservative uncertainties of $\pm\,0.2$~dex in
oxygen abundance and $\pm\,0.3$~mag in galaxy luminosity.
\label{fig-zmb}} 
 
\figcaption[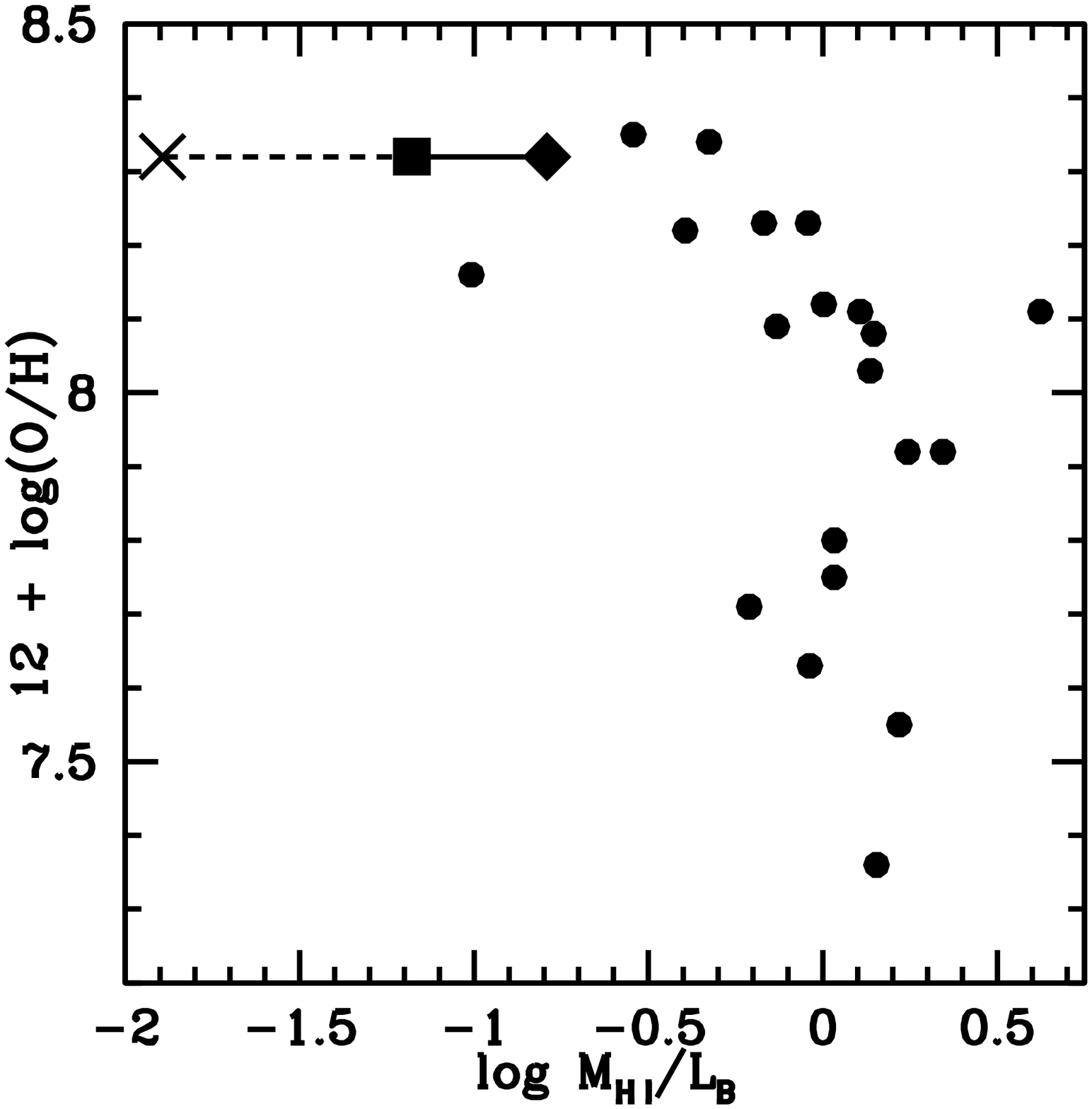]{
Oxygen abundance versus the ratio of the H~I gas mass to blue
luminosity (solar units) for dwarf irregular galaxies.
The filled circles mark the field dwarfs (Richer \& McCall 1995).
The cross shows the position of UGC~7636 based on the H~I flux 
actually measured within the galaxy (Hoffman et al. 1987) in
combination with the oxygen abundance from the newly discovered H~II
region, LR1, inside the H~I cloud STE1. 
The dotted line shows how the location of UGC~7636 would move if the
H~I mass of STE1 were added to the galaxy.
The filled square and the filled diamond mark recent single--dish
measurements of the H~I mass of STE1 by Sancisi, Thonnard, \& Ekers (1987)
and Patterson \& Thuan (1992), respectively.
The last two symbols are connected with a solid line to show the
allowed range for UGC~7636. 
\label{fig-zmh1lb}}
 
\clearpage	% TABLES

\begin{deluxetable}{ccc}
\footnotesize		% small 11pt; footnotesize 10pt; scriptsize 8pt
\tablecaption{Basic data for UGC~7636, STE1, and LR1. 
\label{table-u7636}}
\tablewidth{0pt}
\tablehead{
\colhead{Property} & \colhead{Value} & \colhead{References}
}
\startdata
\cutinhead{{\bf UGC 7636}}
$\alpha$, $\delta$ (J2000) & 
12$^h$30$^m$01\fs0, +07\degr55\arcmin46\arcsec & 1 \\
Morphological type & Im III-IV & 1 \\
$B^0_T$ & 14.26 & 7 \\
$(B-I)^0_T$ & 1.69 & 7 \\
Heliocentric velocity (km s$^{-1}$) & 276 $\pm$ 78 & 2 \\
H I flux integral (Jy km s$^{-1}$) & 0.167 & 3 \\
\cutinhead{{\bf H I cloud STE1}}
$\alpha$, $\delta$ (J2000) & 
12$^h$29$^m$55\fs5, +07\degr57\arcmin38\arcsec & 5 \\
Heliocentric velocity (km s$^{-1}$) & 469 $\pm$ 3 & 5 \\
H I flux integral (Jy km s$^{-1}$) & $1.947 \pm 0.358$ & 6 \\
\cutinhead{{\bf H II region LR1}}
$\alpha$, $\delta$ (J2000) & 
12$^h$29$^m$56$^s$, +07\degr57\arcmin35\arcsec & 4,5,8\tablenotemark{a} \\
Heliocentric velocity (km s$^{-1}$) & $577 \pm 91$ & 8 \\
\enddata
\tablenotetext{a}{
Positional error estimates: 
$\Delta \alpha \simeq \pm 1^s$, $\Delta \delta \simeq \pm 4$\arcsec
}
\tablerefs{
(1) Binggeli et al. 1985; 
(2) Binggeli et al. 1993;
(3) Hoffman et al. 1987;
(4) Lee et al. 1997;
(5) McNamara et al. 1994;
(6) Patterson \& Thuan 1992;
(7) Patterson \& Thuan 1996;
(8) this study
}
\end{deluxetable}

\clearpage

\begin{deluxetable}{lrr}
\footnotesize
\tablecaption{
Emission--line data and derived properties for the H~II region LR1. 
\label{table-data}} 
\tablewidth{0pt}
\tablehead{
\colhead{Identification} & \colhead{$F$} & \colhead{$I$}
}
\startdata
$\rm{[O\;II]}\,\lambda$3727 & $693 \pm 203 (166)$ & $645 \pm 197 (154)$ \\
H$\beta$ & 100 $\pm 17$ & $100 \pm 19$ \\
$\rm{[O\;III]}\,\lambda$4959 & $84 \pm 22 (17)$ & $79 \pm 22 (16)$ \\
$\rm{[O\;III]}\,\lambda$5007 & $130 \pm 31 (22)$ & $121 \pm 31 (21)$ \\
H$\alpha$ & $289 \pm 51 (16)$ & $270 \pm 54 (16)$ \\
$\rm{[N\;II]}\,\lambda$6583 & $13 \pm 13 (13)$ & $12 \pm 12 (12)$ \\
\tableline
Flux$(\rm{H}\beta)$ $(\times 10^{-16}$ ergs s$^{-1}$ cm$^{-2})$ &
	  $3.42 \pm 0.58$ & $3.68 \pm 0.70$ \\
EW(H$\beta$) (\AA) & $26.8 \pm 5.9$ & \\
\tableline
$L(\rm{H}\alpha)$ (ergs s$^{-1}$)\tablenotemark{a} & & $3.3 \times 10^{37}$ \\
SFR ($M_{\odot}$ yr$^{-1}$)\tablenotemark{b} & & $2.9 \times 10^{-4}$ \\
$Q(H^0)$ (s$^{-1}$)\tablenotemark{b} & & $2.4 \times 10^{49}$ \\
12 + log(O/H) & & $8.32 \pm 0.20$ \\
\enddata
\tablenotetext{a}{
We adopt a value of 31.12 mag for the Virgo Cluster distance modulus.
}
\tablenotetext{b}{
The present star formation rate is computed as 
SFR = $L(\rm{H}\alpha)/1.12 \times 10^{41} \; {\rm ergs} \; {\rm s}^{-1}$
(Kennicutt 1983) and
% }
% \tablenotetext{c}{
the number of ionizing photons per second is computed as
$Q(H^0) = 7.43 \times 10^{11} \, L(\rm{H}\alpha)$.
}
\tablecomments{
$F$ is the observed flux ratio with respect to \hbeta\ and
$I$ is the flux ratio corrected for underlying Balmer absorption 
(zero reddening assumed).  
The errors in the line ratios including and excluding (in parentheses)
the error in \hbeta\ flux are listed. 
}
\end{deluxetable}

\clearpage	% INCLUDE FIGURES TO CHECK
 
\begin{figure}
\figurenum{1}
\plotone{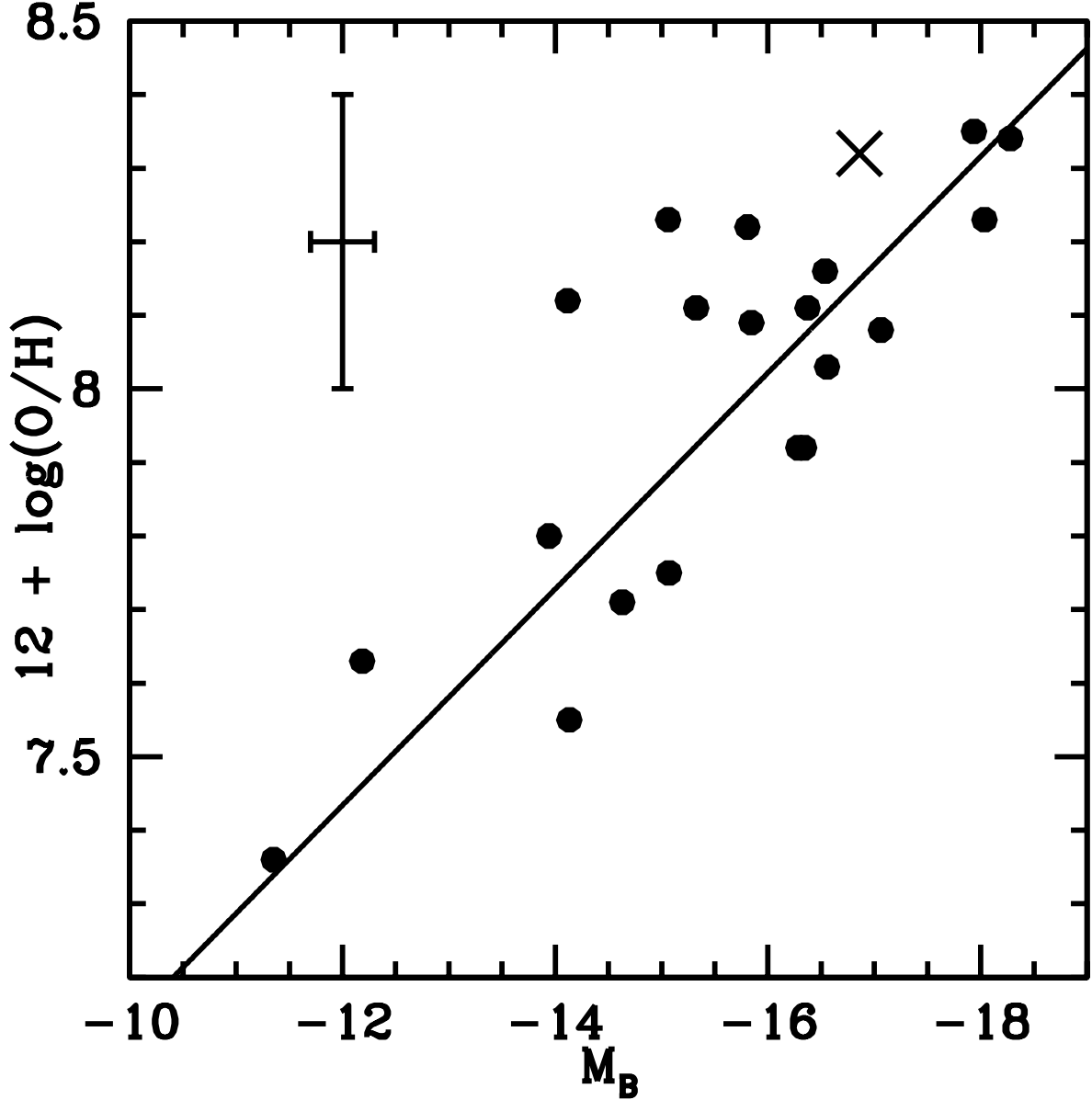}
\caption{
% CAPTION FOR FIGURE 1
Oxygen abundance versus total blue luminosity for dwarf irregular
galaxies.  
The filled circles mark the field dwarfs of Richer \& McCall (1995)
and the solid line is a linear least--squares fit for $M_B < -15$. 
The solar value of the oxygen abundance is 12+log(O/H) = 8.87
(Grevesse, Noels, \& Sauval 1996).
The cross shows where UGC~7636 would lie if its oxygen abundance 
equals that of the newly discovered H~II region, LR1, inside the H~I
cloud STE1. 
The error bar indicates conservative uncertainties of $\pm\,0.2$~dex in
oxygen abundance and $\pm\,0.3$~mag in galaxy luminosity.
}
\end{figure}

\begin{figure}
\figurenum{2}
\plotone{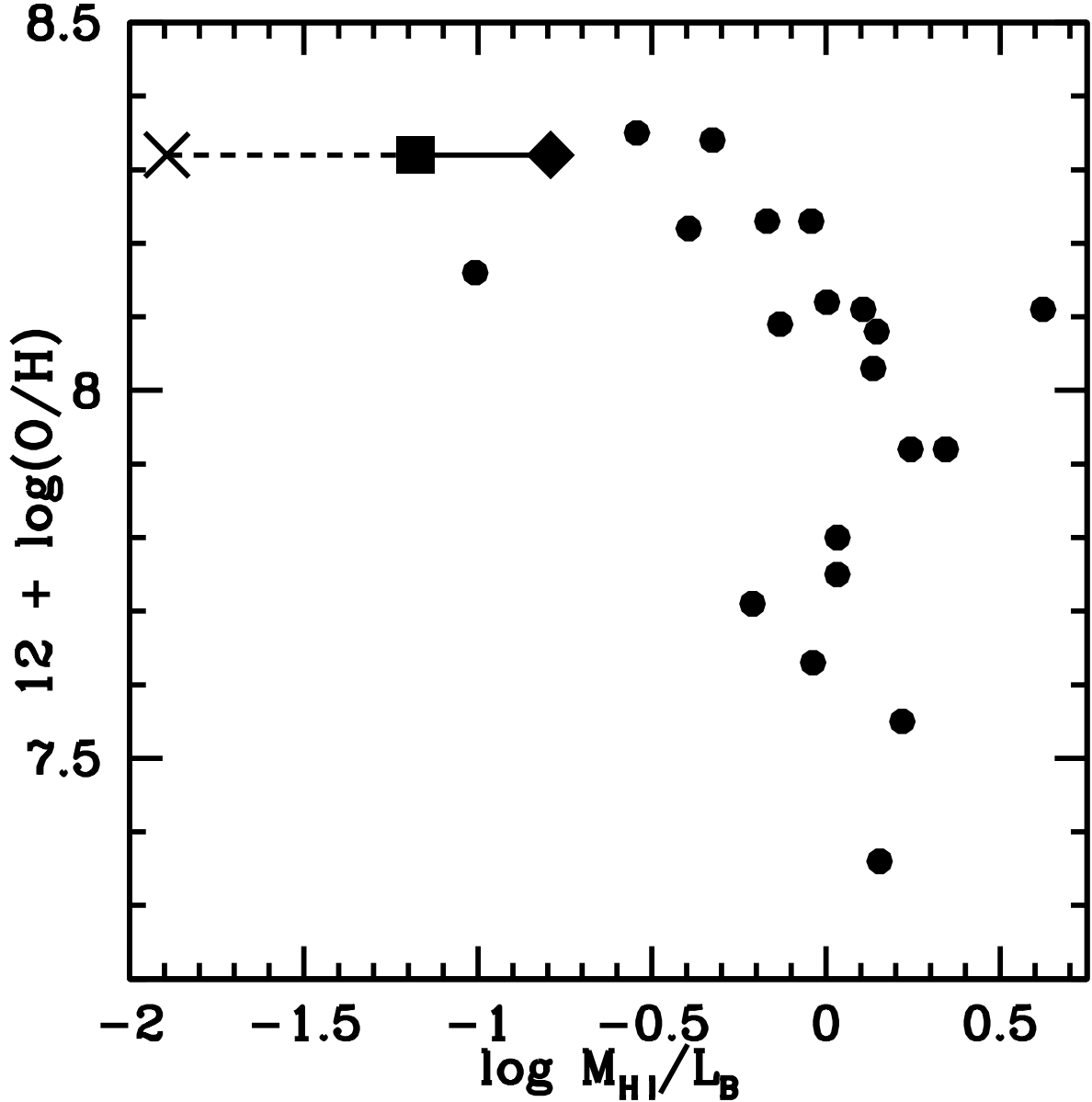}
\caption{
% CAPTION FOR FIGURE 2
Oxygen abundance versus the ratio of the H~I gas mass to blue
luminosity (solar units) for dwarf irregular galaxies.
The filled circles mark the field dwarfs (Richer \& McCall 1995).
The cross shows the position of UGC~7636 based on the H~I flux 
actually measured within the galaxy (Hoffman et al. 1987) in
combination with the oxygen abundance from the newly discovered H~II
region, LR1, inside the H~I cloud STE1. 
The dotted line shows how the location of UGC~7636 would move if the
H~I mass of STE1 were added to the galaxy.
The filled square and the filled diamond mark recent single--dish
measurements of the H~I mass of STE1 by Sancisi, Thonnard, \& Ekers (1987)
and Patterson \& Thuan (1992), respectively.
The last two symbols are connected with a solid line to show the
allowed range for UGC~7636. 
}
\end{figure}

\clearpage

\end{document}